\documentclass[a4paper]{article}

\usepackage{authblk}
\usepackage{graphicx}
\usepackage{amssymb}
\usepackage{appendix}

\newenvironment{mat}[1]{\left[\begin{array}{#1}}{\end{array}\right]}
\newcommand{\bmx}[1]{\begin{mat}{#1}}
\newcommand{\emx}{\end{mat}}

\newcommand{\gssa}[4]{\mbox{\boldmath $#1$}_{#2}^{#3}(#4)}
\newcommand{\ga}[2]{\mbox{\boldmath $#1$}(#2)}
\newcommand{\gss}[3]{\mbox{\boldmath $#1$}_{#2}^{#3}}
\newcommand{\g}[1]{\mbox{\boldmath $#1$}}

\newcommand{\pref}[1]{(\ref{#1})}

\begin{document}

\title{Preamble-based Channel Estimation in OFDM/OQAM Systems: A Review}

\author[1]{Eleftherios Kofidis\thanks{kofidis@unipi.gr}}
\author[2]{Dimitrios Katselis\thanks{dimitrik@kth.se}}
\author[3]{Athanasios~Rontogiannis\thanks{tronto@noa.gr}}
\author[4]{Sergios Theodoridis\thanks{stheodor@di.uoa.gr}}
\affil[1]{Department of Statistics and Insurance Science, University of Piraeus, 80, Karaoli \& Dimitriou str., 185~34 Piraeus, Greece}
\affil[2]{Signal Processing and Automatic Control Laboratories, School of Electrical Engineering, KTH Royal Institute of Technology, Stockholm, Sweden}
\affil[3]{Institute for Astronomy, Astrophysics, Space Applications and Remote Sensing, National Observatory of Athens, Metaxa \& Vas.~Pavlou str., 152~36 Athens, Greece}
\affil[4]{Department of Informatics and Telecommunications, University of Athens, Panepistimioupolis, Ilissia, 157~84 Athens, Greece}

\date{}

\maketitle

\begin{abstract}
Filter bank-based multicarrier communications (FBMC) have recently attracted increased interest in both wired (e.g., xDSL, PLC) and wireless (e.g., cognitive radio) applications, due to their 
enhanced flexibility, higher spectral efficiency, and better spectral containment compared to conventional OFDM. A particular type of FBMC, the so-called FBMC/OQAM or OFDM/OQAM system, consisting of 
pulse shaped OFDM carrying offset QAM (OQAM) symbols, has received increasing attention due to, among other features, its higher spectral efficiency and implementation simplicity. 
It suffers, however, from an imaginary inter-carrier/inter-symbol interference that complicates signal processing tasks such as channel estimation. This paper focuses on channel estimation for OFDM/OQAM
systems based on a known preamble. A review of the existing preamble structures and associated channel estimation methods is given, for both single- (SISO) and multiple-antenna (MIMO) systems. The various preambles are compared via simulations in both mildly and highly frequency selective channels.
\end{abstract}

\section{Introduction}
\label{sec:intro}

Orthogonal frequency division multiplexing (OFDM) has become quite popular in both wired and wireless communications \cite{agm07,hywll09,w09}, mainly because of its immunity to the frequency selectivity of the channel, which allows a significant increase in the transmission rate \cite{np00,lp01}. Using the cyclic prefix (CP) as a guard interval against intersymbol interference (ISI), OFDM manages to turn a frequency selective channel into a set of parallel flat channels with independent noises. This greatly simplifies both the estimation of the channel and the recovery of the transmitted data at the receiver. However, these advantages come at the cost of an increased sensitivity to frequency offset and Doppler spread. This is due to the fact that, although the subcarrier functions are perfectly time limited, they suffer from spectral leakage in the frequency domain and hence inter-carrier interference (ICI) results
under nonideal conditions. Moreover, the inclusion of the CP entails a waste in transmitted power as well as in spectral efficiency, which, in practical systems, can go up to 25\% \cite{agm07}.

An alternative to CP-OFDM, that can mitigate these drawbacks, is given by filter bank-based multicarrier (FBMC) systems, which have recently attracted increased interest \cite{rsfb10} for both wired (e.g., power line comms (PLC) \cite{bm09,brir10,als11}) and wireless (e.g., cognitive radio~\cite{f08,zrrs10,kh11} and DVB-T~\cite{ar11}) applications. For an excellent presentation of FBMC systems, including a review of recent application areas and related standardization activities, the reader could refer to~\cite{f11}. FBMC enjoys enhanced flexibility, higher spectral efficiency, and better spectral containment compared to CP-OFDM~\cite{adlc98,k10}. Special attention has been recently paid to an OFDM filter bank-based variant employing offset quadrature amplitude modulation (OQAM), known as OFDM/OQAM \cite{fab95,f11,j04,jww04}. This was, in its principle, first proposed back in the 60's~\cite{c66,s67}, but the interest in this scheme was only recently revived due to its desirable features including the potential for maximum spectral efficiency as well as implementation simplicity over alternative FBMC schemes~\cite{b10a,b10b,s10}. It consists of pulse shaping, implemented via an IFFT/FFT-based efficient filter bank, and the subcarriers are loaded with staggered OQAM symbols, i.e., real symbols at twice the symbol rate of OFDM/QAM \cite{h81,ssl02}. This allows the pulses to be well localized in both the time and the frequency domains~\cite{b03}, thus increasing the system's robustness to frequency offsets~\cite{bn10,swbsf11} and Doppler effects~\cite{ssb05,mj05,ds07a,ksm08} and providing better spectral containment in bandwidth sensitive applications~\cite{akf06,akpf07,sscd08}. Furthermore, although the OQAM-based OFDM scheme can be seen to exhibit similar peak-to-average power ratio (PAPR) performance with classical OFDM/QAM~\cite{kh12}, the presence of spectral leakage in the latter may ultimately generate higher peak power values~\cite{wbn06}. Moreover, because of the good time localization, a CP does not have to be included in the OFDM/OQAM transmission,\footnote{Nevertheless, this advantage was partly given up in~\cite{ls08a} and a CP-based OFDM/OQAM system was proposed for the sake of facilitating the data reception process. See also~\cite{rvl12}.} which leads to still higher transmission rates~\cite{ssl02,av08,mj05}.

However, the subcarrier functions are now only orthogonal in the \emph{real} field, which means that there is always an \emph{intrinsic} imaginary interference among (neighboring) subcarriers and symbols~\cite{jlr03}. This makes the channel estimation task for OFDM/OQAM systems more challenging. A number of training schemes and associated estimation methods have recently appeared in the literature, including both preamble-based~\cite{kc07,ljlss08,lsl08,hwl09,hwlxl10,dhcl10,kk11a,kk11b} and scattered pilots-based~\cite{jlr03,myok06,kc07,lls08,jj08a,jj08b,yihc08,lls09,bs10,hs10,lsk10,l12} ones. These can be roughly characterized as aiming at cancelling/avoiding the undesired interference or constructively exploiting it to improve the estimation performance. 
The focus of this paper is on OFDM/OQAM channel estimation based on a known preamble. Existing preamble structures and their associated estimation methods are reviewed, for both the single-antenna (SISO) and multiple-antenna (MIMO) scenarios.
The problem of preamble optimization to achieve minimum estimation error is also shortly discussed, based on recently reported results. The estimation performances of the principal methods are compared via simulations for both mildly and highly frequency selective channels. 

The rest of the paper is organized as follows: In Section \ref{sec:model}, the discrete-time baseband equivalent model for the
OFDM/OQAM system is described. Section~\ref{sec:preamblesSISO} reviews the main approaches proposed in the literature for structuring the preamble and exploiting it for SISO channel estimation. The MIMO case is addressed in Section~\ref{sec:preamblesMIMO}.
Comparative simulation results are reported in Section~\ref{sec:sims}. Section~\ref{sec:concls} concludes the paper. 

\noindent {\em Notation.} The complex conjugate of a complex number $z$ is denoted by $z^{\star}$. Also,
$\jmath=\sqrt{-1}$. $\Re(z)$ or $z^{\mathrm{R}}$ denotes the real part of $z$, while its imaginary part is written as $\Im(z)$ or $z^{\mathrm{I}}$. The set of real numbers is denoted by $\mathbb{R}$. 
Boldface lower and upper case letters are used for vectors and matrices, respectively. The $n$th-order identity matrix is denoted by $\gss{I}{n}{}$. 
$\|\cdot\|$ is the Frobenius norm. The expectation is 
denoted by $E(\cdot)$. $\otimes$ is the (left) Kronecker product. 

\section{System Model}
\label{sec:model}

The (QAM or OQAM based) OFDM modulator output is transmitted through a channel of length $L_{h}$, which is, as usual in block transmissions, assumed to be invariant in the duration of an OFDM symbol. 
At the receiver front-end, noise is added, which is assumed Gaussian with zero mean and variance $\sigma^{2}$. For CP-OFDM, the classical configuration is assumed \cite{np00}. 
The discrete-time signal at the
output of an OFDM/OQAM synthesis filter bank (SFB) is given by \cite{ssl02}\footnote{Alternative configurations exist, generating a \emph{real} OFDM/OQAM signal; see \cite{h81,vl01,ls08b}.}
\begin{equation}
s(l)=\sum_{m=0}^{M-1}\sum_{n}d_{m,n}g_{m,n}(l), 
\label{eq:OQAMtxSig}
\end{equation}
where $d_{m,n}$ are \emph{real} OQAM symbols, and
\[
g_{m,n}(l)=g\left(l-n\frac{M}{2}\right)e^{\jmath\frac{2\pi}{M}m\left(l-\frac{L_{g}-1}{2}\right)}e^{\jmath\varphi_{m,n}},
\]
with $g$ being the \emph{real symmetric} prototype filter impulse response (assumed here of unit energy) of length $L_{g}$, $M$ being the \emph{even} number of subcarriers,
and $\varphi_{m,n}=\varphi_{0}+\frac{\pi}{2}(m+n)\mathrm{\ mod\ }\pi$, where $\varphi_{0}$ can be arbitrarily chosen~\cite{ssl02}. In this paper, $\varphi_{m,n}$ is defined as
$(m+n)\frac{\pi}{2}-mn\pi$ as in~\cite{ssl02}. The filter $g$ is usually designed to have length $L_{g}=KM$, with $K$ being the overlapping factor. The double subscript $(\cdot)_{m,n}$ denotes
the $(m,n)$-th frequency-time (FT) point. Thus, $m$ is the subcarrier index and $n$ the OFDM/OQAM symbol time index.

The pulse $g$ is designed so that the associated subcarrier functions $g_{m,n}$ are orthogonal in the \emph{real} field \cite{b03}, that is
\begin{equation}
\Re\left\{\sum_{l}g_{m,n}(l)g^{\star}_{p,q}(l)\right\}=\delta_{m,p}\delta_{n,q},
\label{eq:OQAMorth}
\end{equation}
where $\delta_{i,j}$ is the Kronecker delta. 
This implies that even in the absence of channel distortion and noise,
and with perfect time and frequency synchronization, there will be some intercarrier (and/or intersymbol) interference at the output of the analysis filter bank (AFB), which is 
purely imaginary (the notation in \cite{ljlss08} is adopted): 
\begin{equation}
\sum_{l}g_{m,n}(l)g^{\star}_{p,q}(l)=\jmath \langle g \rangle_{m,n}^{p,q},
\label{eq:innerP}
\end{equation}
and known as \emph{intrinsic} interference \cite{jlr03}. Making the common assumption that the channel is (approximately) frequency flat at each subcarrier
and constant over the duration of the prototype filter \cite{ljlss08}, which is true\footnote{This is not an accurate subchannel model in case $M$, the number of subcarriers, is not large enough with respect to the frequency selectivity of the channel~\cite{msp12} (see Section~\ref{sec:sims} for such examples). This is also the case in high mobility scenarios, where $M$ should be small, to cope with ICI from frequency dispersion \cite{gmm08a,gmm08b,gmm11}.} for practical values of $L_{h}$ and $L_{g}$ and for well localized $g$'s,
one can express the AFB output at the $p$th subcarrier and $q$th OFDM/OQAM symbol as:
\begin{equation}
y_{p,q} = 
H_{p,q}d_{p,q}+\underbrace{\jmath
\underbrace{\sum_{m=0}^{M-1}\sum_{n}}_{(m,n)\neq (p,q)}H_{m,n}d_{m,n}\langle g \rangle_{m,n}^{p,q}}_{I_{p,q}}+\eta_{p,q}
\label{eq:OQAM}
\end{equation}
where $H_{p,q}$ is the channel frequency response (CFR) at that FT point, and $I_{p,q}$ and $\eta_{p,q}$ are the associated interference and noise components, respectively.
$\eta_{p,q}$ has been shown to be stationary (in fact, also Gaussian with zero mean and variance $\sigma^{2}$) and correlated among subcarriers (see \cite{bcmv93}). 
In the sequel, this correlation will be mostly neglected, as it is this case with well localized AFBs. 

For pulses $g$ that are well localized in both time and frequency, the interference from FT points outside a neighborhood $\Omega_{p,q}$ of $(p,q)$ (excluding $(p,q)$ itself) is negligible. If, moreover, the CFR is almost constant over this neighborhood, one can write \pref{eq:OQAM} as
\begin{equation}
y_{p,q}\approx H_{p,q}c_{p,q}+\eta_{p,q}
\label{eq:OQAMapprox}
\end{equation}
where
\begin{equation}
c_{p,q}=d_{p,q}+\jmath\underbrace{\sum_{(m,n)\in\Omega_{p,q}}d_{m,n}\langle g \rangle_{m,n}^{p,q}}_{u_{p,q}}=d_{p,q}+\jmath u_{p,q}
\label{eq:pseudopilot}
\end{equation}
is the \emph{virtual} transmitted symbol at $(p,q)$, with
\begin{equation}
u_{p,q}=\sum_{(m,n)\in\Omega_{p,q}}d_{m,n}\langle g \rangle_{m,n}^{p,q}
\label{eq:u}
\end{equation}  
being the imaginary part of the interference from the neigboring FT points. 
When known pilots are transmitted at that FT point and its neighborhood $\Omega_{p,q}$, the quantity in~\pref{eq:pseudopilot} can be approximated. This can then serve as a \emph{pseudo-pilot} \cite{ljlss08} to compute an estimate of the CFR at the corresponding FT point, as, for example,
\begin{equation}
\hat{H}_{p,q}=\frac{y_{p,q}}{c_{p,q}}\approx H_{p,q}+\frac{\eta_{p,q}}{c_{p,q}}
\label{eq:H=y/x}
\end{equation}
This observation underlies the so-called Interference Approximation Method (IAM) to be desribed in detail in the next section. 

\begin{figure}
\centering
\includegraphics[width=4in]{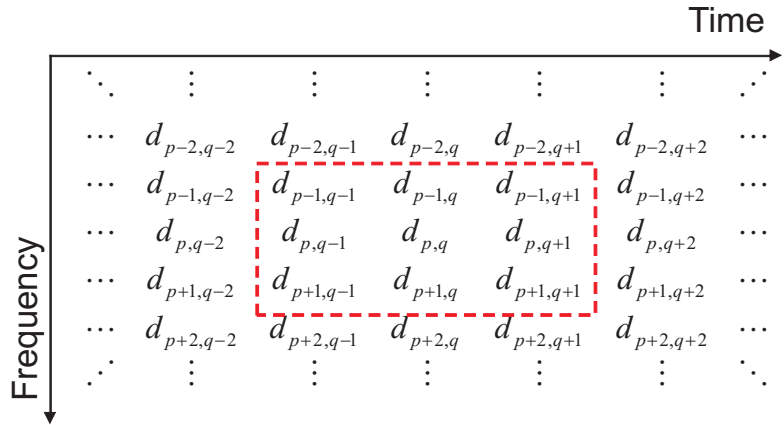}
\caption{The first-order time-frequency neighborhood of FT point $(p,q)$.}
\label{fig:FTneighbors}
\end{figure}

The most common assumption is that, with a well time-frequency localized pulse, contributions to $I_{p,q}$ only come from 
the first-order neighborhood of $(p,q)$, namely
$\Omega_{p,q}=\left\{(p\pm 1,q\pm 1), (p,q \pm 1), (p\pm
1,q)\right\}$ (see Fig.~\ref{fig:FTneighbors}). A special case is given by
$\Omega_{p,q}=\Omega_{p,q}^{1}=\left\{(p\pm 1,q)\right\}$. This
arises when we place three adjacent pilot tones at positions $(p-1,q),(p,q),(p+1,q)$ and
zeros at the rest of the first-order neighborhood positions or when we place nonzero pilot tones at all positions in the training
vector and zero vectors around it. 

If the neighbors of $(p,q)$ carry unknown (data) symbols, one cannot (directly) approximate the imaginary interference in~\pref{eq:pseudopilot}.
However, by properly choosing \emph{one} of the neighboring symbols, say at the point $(r,s)$, this interference can be forced to zero. Then the pseudo-pilot in~\pref{eq:pseudopilot} becomes real, equal to $d_{p,q}$. The pilot at $(r,s)$ is then known as a \emph{help pilot}~\cite{jlr03}.\footnote{This idea was later proposed again in~\cite{kc07}.}

\begin{figure}
\centering
\includegraphics[width=4in]{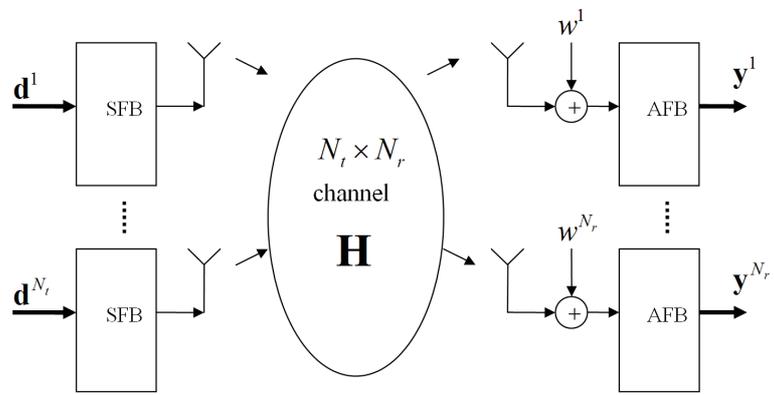}
\caption{The MIMO-OFDM/OQAM system.}
\label{fig:MIMO-FBMC}
\end{figure}

The above formulation can be easily extended to the MIMO case. Consider an $N_{\mathrm{t}}\times N_{\mathrm{r}}$ MIMO system, with identical SFBs (AFBs) at each transmit (receive) antenna (see Fig.~\ref{fig:MIMO-FBMC}).  
Then, adopting the same assumptions as before, one can write an equation analogous to \pref{eq:OQAMapprox} for each receive antenna $j=1,2,\ldots,N_{\mathrm{r}}$:
\begin{equation}
y_{p,q}^{j}\approx \sum_{i=1}^{N_{\mathrm{t}}}H_{p,q}^{j,i}c_{p,q}^{i}+\eta_{p,q}^{j},
\end{equation}
where $H_{p,q}^{j,i}$ is the $M$-point CFR of the channel from the $i$th transmit antenna to the $j$th receive antenna, $c_{p,q}^{i}$ is the corresponding virtual symbol, namely 
\[
c_{p,q}^{i}=d_{p,q}^{i}+\jmath \underbrace{\sum_{(m,n)\in\Omega_{p,q}}d_{m,n}^{i}\langle g \rangle_{m,n}^{p,q}}_{u_{p,q}^{i}}=d_{p,q}^{i}+\jmath u_{p,q}^{i},
\]
and $\eta_{p,q}^{j}$ denotes the corresponding noise component. Clearly, the latter is uncorrelated among different receive antennas. Collecting the above equations for all receive antennas, results, for each FT point, in an input-output equation similar to that for MIMO-OFDM \cite{sbmlip04}:
\begin{equation}
\gss{y}{p,q}{} \approx \gss{H}{p,q}{}\gss{c}{p,q}{}+\gss{\eta}{p,q}{},
\label{eq:OQAMapproxMIMO}
\end{equation}
where
\[
\gss{H}{p,q}{}=\bmx{cccc} H_{p,q}^{1,1} & H_{p,q}^{1,2} & \cdots & H_{p,q}^{1,N_{\mathrm{t}}} \\ H_{p,q}^{2,1} & H_{p,q}^{2,2} & \cdots & H_{p,q}^{2,N_{\mathrm{t}}} \\ 
\vdots & \vdots & \ddots & \vdots \\ H_{p,q}^{N_{\mathrm{r}},1} & H_{p,q}^{N_{\mathrm{r}},2} & \cdots & H_{p,q}^{N_{\mathrm{r}},N_{\mathrm{t}}}\emx
\]
is the MIMO CFR at that point and 
\[
\gss{c}{p,q}{}=\gss{d}{p,q}{}+\jmath \gss{u}{p,q}{}, 
\] 
with 
\[
\gss{d}{p,q}{}=\bmx{cccc} d_{p,q}^{1} & d_{p,q}^{2} & \cdots & d_{p,q}^{N_{\mathrm{t}}} \emx^{T}
\]
and $\gss{u}{p,q}{}$ defined similarly. 

\section{Preamble-based Channel Estimation Methods: The Single-Antenna Case}
\label{sec:preamblesSISO}

Known (training) symbols for the purpose of channel estimation can be gathered in the beginning of a frame and/or placed at isolated FT points throughout the frame. These are the two principal pilot configurations, known as preamble and scattered pilots, respectively. The focus of this paper is on the design of the preamble\footnote{Same principles apply if the pilot blocks are placed 
in the middle of the frame (midambles).} for OFDM/OQAM systems and its use in estimating the channel.
The difficulty in this task, compared to OFDM/QAM, lies in estimating a \emph{complex} CFR 
in the presence of the \emph{imaginary} intrinsic interference among neighboring FT points. The preamble design and the associated estimation method should take this scenario into account in some appropriate way. 
Three different approaches for preamble-based channel estimation, including their variants, are reviewed in this section, for SISO systems. The first approach, Pairs of Pilots (POP), relies
on simple algebraic relations for the input/output samples in a number of (in practice consecutive) time instants. The second one, Interference Approximation Method (IAM), aims at approximating the
intrinsic imaginary interference from neighboring pilots and hence constructing complex pseudo-pilots as above, to accommodate the complex CFR. A number of variants of IAM are considered. In this context, recent results on designing optimal - in the sense of minimizing the channel estimation error - preambles are also recalled.
An alternative approach, that of cancelling or avoiding interference, is then discussed.  The extension of some of these methods to the MIMO case is addressed in the following section. 

Note that the methods presented here can be applied to any OFDM/OQAM filter bank, provided, of course, that the prototype filter is sufficiently well localized in time and frequency 
and the number of subcarriers is large enough that the models in eqs.~\pref{eq:OQAMapprox}, \pref{eq:OQAMapproxMIMO} hold with sufficient accuracy. 
Thus, methods aiming at enhancing the channel estimation performance through an interference-minimizing filter design are omitted; see, e.g., \cite{dsb06,ls09}.
Moreover, we do not include methods based on preambles that are not completely known to the receiver (as in, e.g., \cite{gvm08}, where the training sequence is superimposed on the data) or
are too spectrally inefficient (e.g., made of isolated pilots surrounded by a large neighborhood of zeros, as in~\cite{myok05,myok06,myk09}) or especially suited to wideband sub-channels,
being significantly more complicated (e.g., \cite{gmm08b}) or significantly longer than those considered here (e.g., \cite{hir08,hs10}). 

In the sequel, and in addition to the assumptions stated in the previous section, the channel is assumed to be time invariant during the preamble period. 

\subsection{Pairs of Pilots (POP) Method}
\label{subsec:POP}

This method, proposed in~\cite{ljlss08}, aims at computing a channel estimate by using~\pref{eq:OQAMapprox} at two different (in practice
consecutive) time instants, $q_{1},q_{2}$, to construct a system of equations for the real and imaginary parts
of $H_{p,q}$. Here, it is described in an \emph{alternative, equivalent way} \cite{d3.1}. 

With the noise being \emph{neglected}, denote by $W_{p,q}=1/H_{p,q}$ the corresponding zero forcing (ZF) equalizer coefficients. Then, one can write 
\begin{eqnarray*}
y_{p,q_{1}}W_{p,q_{1}} & = & d_{p,q_{1}}+\jmath u_{p,q_{1}} \\
y_{p,q_{2}}W_{p,q_{2}} & = & d_{p,q_{2}}+\jmath u_{p,q_{2}}
\end{eqnarray*} 
and therefore 
\begin{eqnarray*}
y_{p,q_{1}}^{\mathrm{R}}W_{p,q_{1}}^{\mathrm{R}}-y_{p,q_{1}}^{\mathrm{I}}W_{p,q_{1}}^{\mathrm{I}} & = & d_{p,q_{1}} \\
y_{p,q_{2}}^{\mathrm{R}}W_{p,q_{2}}^{\mathrm{R}}-y_{p,q_{2}}^{\mathrm{I}}W_{p,q_{2}}^{\mathrm{I}} & = & d_{p,q_{2}}
\end{eqnarray*} 
Noting that $W_{p,q_{1}}\approx W_{p,q_{2}}$ due to the (almost) time invariance of the CFR,
\[
\bmx{cc} y_{p,q_{1}}^{\mathrm{R}} & -y_{p,q_{1}}^{\mathrm{I}} \\ y_{p,q_{2}}^{\mathrm{R}} & -y_{p,q_{2}}^{\mathrm{I}} \emx \bmx{c} W_{p,q_{1}}^{\mathrm{R}} \\ W_{p,q_{1}}^{\mathrm{I}} \emx = \bmx{c} d_{p,q_{1}} \\ d_{p,q_{2}} \emx,
\]
hence
\[
\bmx{c} W_{p,q_{1}}^{\mathrm{R}} \\ W_{p,q_{1}}^{\mathrm{I}} \emx = \frac{1}{y_{p,q_{1}}^{\mathrm{I}}y_{p,q_{2}}^{\mathrm{R}}-y_{p,q_{1}}^{\mathrm{R}}y_{p,q_{2}}^{\mathrm{I}}}
\bmx{c} y_{p,q_{1}}^{\mathrm{I}}d_{p,q_{2}}-y_{p,q_{2}}^{\mathrm{I}}d_{p,q_{1}} \\ y_{p,q_{1}}^{\mathrm{R}}d_{p,q_{2}}-y_{p,q_{2}}^{\mathrm{R}}d_{p,q_{1}} \emx
\]
This can be more compactly written as
\begin{equation}
W_{p,q_{1}}=W_{p,q_{1}}^{\mathrm{R}}+\jmath W_{p,q_{1}}^{\mathrm{I}}=\jmath \frac{d_{p,q_{1}}y_{p,q_{2}}^{\star}-d_{p,q_{2}}y_{p,q_{1}}^{\star}}{\Im(y_{p,q_{1}}^{\star}y_{p,q_{2}})}
\end{equation}
In practice, the preamble would comprise the first two OFDM/OQAM symbols, that is, $q_{1}=0$ and $q_{2}=1$. In that case, and with the preamble suggested in~\cite{ljlss08}, where
$d_{p,0}=(-1)^{p}$ and the second symbol is all zeros, the above would simplify to
\begin{equation}
W_{p,0}=\jmath \frac{(-1)^{p}y_{p,1}^{\star}}{\Im(y_{p,0}^{\star}y_{p,1})}
\label{eq:POP0}
\end{equation}
The CFR would then be computed as $H_{p,0}=1/W_{p,0}$.

An advantage of the POP scheme, besides its simplicity, is that it does not explicitly depend on the
employed prototype filter. However, it must be emphasized that the above derivation only holds
when noise is negligible. One can see that (see \cite{ljlss08}), in the presence of noise, the method can have
unpredictable performance since the degree of the noise enhancement in general also depends on
unknown (hence uncontrollable) data.

\subsection{Interference Approximation Method (IAM)}
\label{subsec:IAMSISO}

If the interference $\jmath u_{p,q}$ in \pref{eq:pseudopilot} is only due to the immediate neighbors of $(p,q)$ and these FT points carry 
training (hence known) symbols, then one can compute an approximation of this interference, as pointed out above. This
can then be used to construct the complex pseudo-pilots $c_{p,q}$ and subsequently use them to get a channel estimate in exactly the same way as in CP-OFDM (see~\pref{eq:H=y/x}).
This method is called (for obvious reasons) the Interference Approximation Method (IAM) and
requires that all input symbols in the immediate neighborhood of $(p,q)$ be known. This implies that, for each subcarrier,
(at least) 1.5 complex symbols are required for training, representing an increased (though not significantly)
overhead compared to CP-OFDM. The training would need to extend over more than three time instants for less well time-localized filters.

Writing \pref{eq:H=y/x} out,
\[
\hat{H}_{p,q}=H_{p,q}+\frac{\eta_{p,q}}{d_{p,q}+\jmath u_{p,q}},
\]
one can see that the pilots should be so chosen as to minimize the effect of the noise term.
In \cite{ljlss07,ljlss08}, it is proposed that these be chosen so as to have pseudo-pilots of maximum magnitude. To this end, the
surrounding training symbols should be such that all terms in~\pref{eq:u} have the same sign so that they
add together. Moreover, this should happen at \emph{all} frequencies $p$. 

To this end, one needs to know the interference weights $\langle g \rangle_{m,n}^{p,q}$ for the neighbors $(m,n)\in \Omega_{p,q}$ of each FT point $(p,q)$ of interest. 
These can be \emph{a priori} computed based on the employed prototype filter $g$. Moreover, it can be shown (see the Appendix) that, for \emph{any} choice of $g$, these weights follow a specific pattern, which, for the definition of $\varphi_{0}$ adopted here,
and for \emph{all} $q$, can be written as\footnote{This is without loss of generality as analogous patterns result for other definitions of $\varphi_{0}$ (see, e.g., \cite{d08}).}
\begin{equation}
\begin{array}{rrr}
(-1)^{p}\epsilon & 0 & -(-1)^{p}\epsilon \\
& & \\
(-1)^{p}\delta & -\beta & (-1)^{p}\delta \\
& & \\
-(-1)^{p}\gamma & d_{p,q} & (-1)^{p}\gamma \\
& & \\
(-1)^{p}\delta & \beta & (-1)^{p}\delta \\
& & \\
(-1)^{p}\epsilon & 0 & -(-1)^{p}\epsilon 
\end{array}
\label{eq:pattern_}
\end{equation}
with the horizontal direction corresponding to time and the vertical one to frequency, as in Fig.~\ref{fig:FTneighbors}.
The above quantities can be shown (see the Appendix) to be given by
\begin{eqnarray}
\beta & = & e^{-j\frac{2\pi}{M}\frac{L_{g}-1}{2}}\sum_{l=0}^{L_{g}-1}g^{2}(l)e^{j\frac{2\pi}{M}l} \\
\gamma & = & \sum_{l=\frac{M}{2}}^{L_{g}-1}g(l)g\left(l-\frac{M}{2}\right) \\
\delta & = & -je^{-j \frac{2\pi}{M}\frac{L_{g}-1}{2}}\sum_{l=M/2}^{L_{g}-1}g(l)g\left(l-\frac{M}{2}\right)e^{j \frac{2\pi}{M}l} \\
\epsilon & = & e^{\mp\jmath \frac{2\pi}{M}(L_{g}-1)}\sum_{l=M/2}^{L_{g}-1}g(l)g\left(l-\frac{M}{2}\right)e^{\pm \jmath 2 \frac{2\pi}{M}l}
\end{eqnarray}
and are (absolutely) smaller than one, with $\beta,\gamma,\delta > 0$.  
Generally, $\beta, \gamma > \delta > |\epsilon|$. For example, $\gamma=0.553$, $\beta=0.25$, $\delta=0.2172$, and $\epsilon\approx 0.0004$ for the $g$'s designed in~\cite{b01} with $M=512,K=3$. 

\subsubsection{The IAM-R Method}

In order to simplify the task of generating pseudo-pilots of large magnitude, it was suggested in~\cite{ljlss07,ljlss08,l08} to place nulls at the first and third OFDM/OQAM symbols of the preamble,
that is, $d_{p,0}=d_{p,2}=0$, $p=0,1,\ldots,M-1$. Then, the imaginary part of the pseudo-pilot at $(p,1)$ will only come from the symbols at the positions $(p\pm 1,1)$. Obviously, these pilots must be OQAM symbols of maximum modulus, $d$. Moreover, in view of the pattern~\pref{eq:pattern_}, they should satisfy the condition $d_{p+1,1}=-d_{p-1,1}$ for all $p$, in order to yield pseudo-pilots of maximum magnitude, namely $|d_{p,1}+\jmath 2\beta d_{p+1,1}|=d\sqrt{1+4\beta^{2}}$. 
This preamble structure is shown in Fig.~\ref{fig:IAMSISO}(a), for the example of $M=8$ and with OQPSK modulation. 
This channel estimation method is called IAM2 in~\cite{ljlss08}, to distinguish it from IAM1, where all symbols, at all three time instants, are randomly chosen. 
To prevent possible increase of the peak to average power ratio (PAPR) with IAM2, a random choice of the middle symbols in IAM2 is also proposed in~\cite{ljlss08}, giving rise to the IAM3 method. 
In the sequel, we will focus on IAM2, due to its ability to maximize the magnitude of the pseudo-pilots with the aid of \emph{real} pilot symbols, and use the name IAM-R, coined in~\cite{lsl08} to signify the fact that its pilot symbols are real.

\subsubsection{The IAM-I Method}

It can be verified from \pref{eq:pseudopilot} that one can do better in maximizing the magnitude of the pseudo-pilots if \emph{imaginary} pilot symbols $\jmath d_{p,1}$ of maximum modulus $d$ are also allowed in the preamble. The reason is that the corresponding pseudo-pilot would then be imaginary and hence of larger magnitude than in IAM-R, namely $|d_{p,1}+u_{p,1}|$, provided that the signs of the pilot symbol and its neighbors are properly chosen so that $d_{p,1}u_{p,1}>0$. Such an IAM scheme was first proposed in~\cite{lsl08}, and was called IAM-I to signify the presence of imaginary pilots. The middle preamble vector consists of triplets, with each one of them following the above principle, and the pilots in each triplet are otherwise selected in random and independently of the other triplets. Hence, imaginary pseudo-pilots result in only one third of the subcarriers, with magnitude $d(1+2\beta)$, whereas the rest of them deliver complex pseudo-pilots of smaller magnitude, $d|(1+\beta)+\jmath\beta|$. 
An example, for $M=8$ and OQPSK symbols, is depicted in Fig.~\ref{fig:IAMSISO}(b). Note that the above preamble is, strictly speaking, not OQAM. Nevertheless, it can still be fed to the
synthesis filter bank and perfectly reconstructed at the analysis bank. 

\subsubsection{The IAM-C Method}

With the preamble proposed in~\cite{lsl08}, imaginary pseudo-pilots will result in only one third of the subcarriers used, whereas the rest of the subcarriers will deliver complex pseudo-pilots. With a slight modification, one can improve upon this and obtain pseudo-pilots that are either purely real or imaginary at \emph{all} the subcarriers. 
The idea is to simply set the middle OFDM/OQAM symbol equal to that in IAM-R but with the pilots at the odd subcarriers multiplied by $\jmath$, as shown in~\cite{d4.1} and independently in~\cite{ds09}. Again, for the $M=8$ OQPSK example, this will result in the preamble shown in Fig.~\ref{fig:IAMSISO}(c). This preamble was called IAM-C in~\cite{ds09}.

\subsubsection{Optimal Preambles}

In~\cite{ds09}, the IAM-C preamble was derived as a solution to the problem of optimal 3-symbol preamble design with zero guard symbols, where optimal is in the sense of generating pseudo-pilots of
maximum magnitude. Optimality of preambles of this kind, in the sense of attaining the minimum possible mean squared error (MSE) with a given transmit energy budget, was recently studied in~\cite{kkrt10} and the results were later complemented in~\cite{kbrhk11}. Two different configurations were considered: \emph{sparse}, where only $L_{h}$ isolated subcarriers carry pilots, with the rest of them being nulled, and
\emph{full}, with all frequencies being occupied with nonzero pilots. It was proved that:
\begin{enumerate}
\item The pilot-carrying subcarriers in an optimal sparse preamble are equidistant with the pilots being equipowered.
\item With $\varphi_{m,n}$ defined as above and including the phase factors $e^{\jmath\varphi_{m,1}}$ in the pilot symbols, they should be all equal but with alternating signs in an optimal full preamble.\footnote{With $\varphi_{m,n}=(m+n)\pi/2$ instead, these symbols must be all equal. See~\cite{kbrhk11} for more on this, independent of the $\varphi_{m,n}$ definition.}
\end{enumerate}
One can readily see that the IAM-C preamble -- proved to be optimal in \cite{ds09} -- follows the general structure of an optimal full preamble of this kind, as it was derived in~\cite{kkrt10,kbrhk11}. For example, multiplied with $e^{j\varphi_{m,1}}$ the middle vector in Fig.~\ref{fig:IAMSISO}(c) becomes $\jmath,-\jmath,\jmath,-\jmath,\ldots$. 

\subsubsection{The Extended IAM-C (E-IAM-C) Method}

The optimality of IAM-C is restricted to the 3-symbol preambles with nulls at their sides.
One could wonder whether an alternative preamble structure which employs the side symbols as well could yield pseudo-pilots that are stronger than those of IAM-C. 
It turns out that such a preamble can indeed be formed in an analogous way as previously, by employing the left- and right-hand sides of the neighborhood~\pref{eq:pattern_} as well~\cite{kk11a}.\footnote{This idea seems to have first been investigated in~\cite[Section~5.4]{d08}. It must be noted however, that, for a number of pulse types (including those of~\cite{b01} employed for the simulations in this paper, where $\gamma > \beta > \delta$ and $\delta < 0.5$), the preamble of~\cite{kk11a} results in pseudo-pilots of a larger magnitude than in the preambles developed in~\cite{d08}.}
Thus, at an odd-indexed subcarrier $p$, with the pilot $\pm d\jmath$ in the middle, one should place $\mp d$ at the right hand side, and its negative, $\pm d$ at the left hand side. 
These side pilots will then contribute $\pm d\gamma \jmath$ each to the interference for the central pilot. In an analogous way, at an even-indexed subcarrier $p$, the middle pilot, say $\pm d$, will get a total interference of 
$\pm 2d\gamma\jmath$ from its $(p,0)$ and $(p,2)$ neighbors, if these are chosen as $\pm d\jmath$ and $\mp d\jmath$, respectively. It can then be verified that,
because of \pref{eq:pattern_}, the interference components with weights $\pm \delta$ cancel each other, and hence have no contribution to the pseudo-pilot. However, this loss is not significant
since $\delta$ is significantly smaller than $\gamma$. The resulting pseudo-pilots are again either purely real or imaginary, with magnitude $d|1+2(\beta+\gamma+2\epsilon)|$, which is clearly larger than that achieved by IAM-C. For the example filters employed here~\cite{b01}, with $M=512$ and $K=3$, the corresponding magnitudes are $2.6076d$ and $1.5d$.
An example of the preamble configuration for $M=8$ and OQPSK modulation is given in Fig.~\ref{fig:IAMSISO}(d). 
One can see that the left hand column is built by repeating each quadruple of the middle pilots in reverse order, while the right hand column is simply the negative of the left hand one. 

The above hold for prototype filters having $\epsilon\geq 0$, as it is most often the case. Whenever $\epsilon<0$, as it is the case for Extended Gaussian Function (EGF)-based filter banks for example~\cite{l08}, the E-IAM-C preamle 
can be modified to yield pseudo-pilots of magnitude $\sqrt{(1+2\beta)^{2}+4(\gamma-2\epsilon)^{2}}$. Fig.~\ref{fig:IAMSISO}(e) shows such a preamble configuration for $M=8$, with OQPSK modulation. 

It must be emphasized here that, since all three OFDM/OQAM symbols of the E-IAM-C preamble are nonzero, it will require more power to be transmitted than the previous ones. This fact is \emph{fairly} taken into account in the simulation experiments (see Section~\ref{sec:sims}), along with possible interference coming from the data part of the frame.

\subsection{Interference Cancellation/Avoidance Methods}
\label{subsec:ICM}

An alternative, more straightforward approach to cope with the intrinsic interference is to so design the preamble and the associated estimation method as to somehow \emph{cancel} or \emph{avoid} it. This is 
in contrast to the IAM approach presented above, where interference is taken advantage of to improve the channel estimation accuracy. The methods following this alternative approach can be roughly categorized as follows.

A \emph{direct} way to cancel the interference is to simply null the data surrounding the pilot FT point. Note that this idea was already used in some of the IAM preambles above, where 
guard OFDM/OQAM symbols were inserted to (approximately) null the interference from preceding and following data. The problem can be further simplified 
if zeros are also transmitted at the middle symbol, at all odd- (or all even-) indexed subcarriers. In this way, the CFR can be estimated at the even-(odd-)indexed subcarriers just like in CP-OFDM.
The channel at the rest of the frequencies will then have to be estimated via interpolation~\cite{hss07}. Such a preamble design was proposed in~\cite{hwlxl10} and an example (for $M=8$) is given in
Fig.~\ref{fig:ICMpreamble}(a). 
The same idea was essentially used in~\cite{myok06,myk09}, and it was also independently included in the preamble evaluation study of~\cite{d4.1} (for MIMO). In the latter, only every third subcarrier was modulated, in accordance with
the pilot configuration suggested in WiMAX DL-PUSC with segmentation \cite{agm07}. A less direct design, which relies on the symmetries in~\pref{eq:pattern_} to cancel the interference from
adjacent subcarriers, was proposed in~\cite{kc07} and is depicted in~Fig.~\ref{fig:ICMpreamble}(b) for the $M=8$ example. 

A more spectrally efficient preamble design method, which also takes advantage of the symmetries in the neighboring times in~\pref{eq:pattern_} to transmit data at these FT positions instead of nulls, was suggested in~\cite{yihc08}. 
The way this is achieved is by considering only the first-order neighbors and canceling the 
interferences coming from them in pairs. Thus, following the pattern in~\pref{eq:pattern_}, the data modulating the first-order neighbors would be structured as follows:
\[
\begin{array}{rrr}
a & b & c \\
d & \times & d \\
-a & b & -c
\end{array}
\]
Note that the idea is similar to that of the help pilot~\cite{jlr03}, although, here, only four, instead of seven information symbols can be transmitted per pilot. 
In fact, this idea is proposed in~\cite{yihc08} in the context of scattered pilots. One can however easily use it to design a preamble with the property of interference cancellation, by simply
applying the above pattern for every third subcarrier and resorting to interpolation for the rest of them. This would result in a preamble containing about $\frac{4M}{3}$ data symbols; this is a considerable 
gain in spectral efficiency compared to the methods previously described. To obtain better channel estimates at the intermediate frequencies, one could even apply this in every second subcarrier. 
This is made possible by the fact that, as shown in~\pref{eq:pattern_}, the interference weights are the same for every second subcarrier. 
However, such a preamble construction would impose some additional constraints to the data that can be transmitted, reducing the data symbols to only $3+\frac{M}{2}$. Fig.~\ref{fig:ICMpreamble}(c) provides an example for $M=8$ and pilots placed at the even-indexed subcarriers.  

Preambles of an even higher spectral efficiency can be conceived, comprising only \emph{one} reference OFDM/OQAM symbol surrounded by \emph{completely unknown} data. Hence, no guard symbols or data of a specific structure are transmitted along with the known symbols as in the previous cases. This, however, implies that the interference from surrounding symbols is completely unknown and its elimination becomes an important issue. 
The approach proposed in~\cite{hwl09} is to follow an iterative procedure, where in each iteration the available channel estimate is used to approximate the surrounding data and vice versa. The original channel estimate is only based on the preamble symbol, ignoring the interference effect. The following channel estimates are computed on the basis of both pilot symbols and estimated data symbols (e.g., using least squares (LS)). The channel and data estimates are improved from one iteration to the other. In the simulation results reported in~\cite{hwl09}, the iterative approach is shown to achieve better performance than the method relying on guard symbols, in only two iterations. More iterations are shown to lead to further improvements. 
Such an iterative approach was also proposed in~\cite{lls09}, albeit for isolated (scattered) pilots. 
Observe that one can so choose the pilot symbols so as to, at least, cancel the interference from neighboring pilots, by respecting the symmetry of~\pref{eq:pattern_}. An example, for $M=8$, is given in Fig.~\ref{fig:ICMpreamble}(d). 
A similar preamble is used in the method of~\cite{dhcl10}, where the channel and data symbol estimates are used to approximate the interference and subtract it from the AFB output. 

\section{Preamble-based Channel Estimation Methods: The Multiple-Antenna Case}
\label{sec:preamblesMIMO}

\subsection{MIMO POP Method}
\label{sec:POPMIMO}

The POP method of Section~\ref{subsec:POP} can be easily extended to the MIMO case, as follows.
Consider \pref{eq:OQAMapproxMIMO} with noise being neglected and assume $\gss{H}{p,q}{}$ is left-invertible for all $p,q$ (hence $N_{\mathrm{r}}\geq N_{\mathrm{t}}$). 
Denote by $\g{W}$ the (per-subcarrier) ZF equalization matrix, that is, $\gss{W}{p,q}{}=\gss{H}{p,q}{\dag}$, where $\gss{H}{p,q}{\dag}$ is the left inverse of $\gss{H}{p,q}{}$.
Write~\pref{eq:OQAMapproxMIMO} in the form $\gss{W}{p,q}{}\gss{y}{p,q}{}=\gss{c}{p,q}{}$ for $2N_{\mathrm{r}}$ time instants, say $q=0,1,\ldots,2N_{\mathrm{r}}-1$,
assuming that the channel (and hence $\g{W}$) can be viewed as invariant in this time period. Then
\[
\gss{W}{p,0}{\mathrm{R}}\gss{y}{p,q}{\mathrm{R}}-\gss{W}{p,0}{\mathrm{I}}\gss{y}{p,q}{\mathrm{I}}=\gss{d}{p,q}{},
\]
for $q=0,1,\ldots,2N_{\mathrm{r}}-1$. Collecting these equations into one, one can write
\[
\bmx{cc} \gss{W}{p,0}{\mathrm{R}} & \gss{W}{p,0}{\mathrm{I}} \emx 
\underbrace{\bmx{cccc} \gss{y}{p,0}{\mathrm{R}} & \gss{y}{p,1}{\mathrm{R}} & \cdots & \gss{y}{p,2N_{\mathrm{r}}-1}{\mathrm{R}} \\
           -\gss{y}{p,0}{\mathrm{I}} & -\gss{y}{p,1}{\mathrm{I}} & \cdots & -\gss{y}{p,2N_{\mathrm{r}}-1}{\mathrm{I}}
\emx}_{\gss{Y}{p}{}} = 
\underbrace{\bmx{cccc} \gss{d}{p,0}{} & \gss{d}{p,1}{} & \cdots & \gss{d}{p,2N_{\mathrm{r}}-1}{} \emx}_{\gss{D}{p}{}}
\]
or
\[
\bmx{cc} \gss{W}{p,0}{\mathrm{R}} & \gss{W}{p,0}{\mathrm{I}} \emx \gss{Y}{p}{} = \gss{D}{p}{}
\]
The ZF equalizer can then be computed as
\[
\bmx{cc} \gss{W}{p,0}{\mathrm{R}} & \gss{W}{p,0}{\mathrm{I}} \emx = \gss{D}{p}{}\gss{Y}{p}{-1}
\]
and hence the channel matrix can be found as the \emph{right inverse} of 
\[
\gss{W}{p,0}{}=\gss{W}{p,0}{\mathrm{R}}+\jmath\gss{W}{p,0}{\mathrm{I}}=\gss{D}{p}{}\gss{Y}{p}{-1}\left(\bmx{c} 1 \\ \jmath \emx \otimes \gss{I}{N_{\mathrm{r}}}{}\right)
\]

\noindent
\emph{Remark.} It must be noted that this scheme is no longer associated to the concept of ``pairs of pilots" and the name ``POP" can only be used here as a reminder of its SISO counterpart. 

\subsection{MIMO IAM}
\label{subsec:IAMMIMO}

As in MIMO-OFDM, one can consider three different patterns for preamble training (see, e.g.,~\cite[Section~5.7]{agm07}): independent (one antenna at a time), scattered (transmitting at different
frequencies), and orthogonal (transmitting orthogonal signals at the same times and
frequencies). IAM preambles of the latter category can be easily constructed from their SISO counterparts if a pattern analogous to that used for MIMO-OFDM in~\cite{sbmlip04} is followed. 
Such an extension of IAM to the MIMO setup was reported and analyzed in~\cite{kk11b}, and is, in fact, the first attempt to address \emph{full} preamble-based channel estimation in the MIMO-OFDM/OQAM context. A \emph{sparse} preamble design was first considered in~\cite{d4.1} and was based on earlier work on \emph{orthogonal} MSE-optimal training design for MIMO-OFDM systems. This will be described in the next subsection. 

It is clear from~\pref{eq:OQAMapproxMIMO} that one will need at least $N_{\mathrm{t}}$ (nonzero) OFDM/OQAM symbols to estimate the CFR matrix. Assume the channel is time invariant 
during a period of $2N_{\mathrm{t}}+1$ OFDM/OQAM symbols ($3N_{\mathrm{t}}$ for E-IAM-C). Take the example of $N_{\mathrm{t}}=2$, for the sake of simplicity. One antenna can use the same preamble as the one used in the SISO case, but with a repetition, as shown in the example of~Fig.~\ref{fig:IAMMIMO}(a), for IAM-C (similarly for the other IAM variants).
The other antenna uses the same preamble, 
but with reversed signs at the second nonzero OFDM/OQAM symbol (cf.\ Fig.~\ref{fig:IAMMIMO}(b)). Then, writing~\pref{eq:OQAMapproxMIMO} at times $q=1,3$ ($q=1,4$ for E-IAM-C) results in
\[
\bmx{cc} \gss{y}{p,1}{} & \gss{y}{p,3}{} \emx = \gss{H}{p,1}{}\bmx{cc} c^{1}_{p,1} & c^{1}_{p,3} \\ c^{2}_{p,1} & c^{2}_{p,3}\emx + \bmx{cc} \gss{\eta}{p,1}{} & \gss{\eta}{p,3}{} \emx
\]
Taking the structure of this preamble into account and recalling the assumption about the interference being mostly contributed from the first-order FT neighbors, one can easily see that
$c^{1}_{p,1}=c^{1}_{p,3}=c^{2}_{p,1}=-c^{2}_{p,3}\equiv c_{p}$ and hence 
\begin{eqnarray}
\bmx{cc} \gss{y}{p,1}{} & \gss{y}{p,3}{} \emx & = & \gss{H}{p,1}{}\bmx{cr} c_{p} & c_{p} \\ c_{p} & -c_{p} \emx + \bmx{cc} \gss{\eta}{p,1}{} & \gss{\eta}{p,3}{} \emx \nonumber \\
& = & \gss{H}{p,1}{}c_{p}\gss{A}{2}{} + \bmx{cc} \gss{\eta}{p,1}{} & \gss{\eta}{p,3}{} \emx, \label{eq:y1y3}
\end{eqnarray}
where $\gss{A}{2}{}$ is the \emph{orthogonal} matrix 
\[
\gss{A}{2}{}=\bmx{cr} 1 & 1 \\ 1 & -1 \emx
\]
and the pseudo-pilot $c_{p}$ can be again pre-calculated as in the SISO case. An estimate of the CFR matrix at the subcarrier $p$ can then be computed as
\begin{eqnarray}
\gss{\hat{H}}{p,1}{} & = & \bmx{cc} \gss{y}{p,1}{} & \gss{y}{p,3}{} \emx\frac{1}{c_{p}}\gss{A}{2}{-1} \nonumber \\
& = & \gss{H}{p,1}{} + \frac{1}{2c_{p}}\bmx{cc} \gss{\eta}{p,1}{} & \gss{\eta}{p,3}{} \emx \gss{A}{2}{} \label{eq:HhatMIMO}
\end{eqnarray}

\noindent
\emph{Remarks.}
\begin{enumerate}
\item It should be noted that, in all IAM variants except for IAM-I, $c_{p}$'s have the same magnitude 
at all frequencies $p$.
\item Note that, in view of the orthogonality of the matrix $\gss{A}{2}{}$, noise enhancement is again controlled by the magnitude of the pseudo pilot $c_{p}$. Indeed, one can readily see that the covariance matrix of the noise term in~\pref{eq:HhatMIMO} equals $\frac{\sigma^{2}}{|c_{p}|^{2}}\gss{I}{N_{\mathrm{r}}}{}$, which corresponds to an MSE given by
\begin{equation}
E\left(\|\gss{H}{p,1}{}-\gss{\hat{H}}{p,1}{}\|^{2}\right)=N_{\mathrm{r}}\sigma^{2}/|c_{p}|^{2}
\label{eq:MSE}
\end{equation}
\item The preamble given above is easily generalized to systems with more than two transmit antennas, with $N_{\mathrm{t}}$ being a power of~2. Simply select the corresponding matrix $\gss{A}{N_{\mathrm{t}}}{}$ to be a Hadamard matrix of order $N_{\mathrm{t}}$ and build the preamble accordingly. The MSE will be still given by~\pref{eq:MSE}.
\item For an $N_{\mathrm{t}}\times N_{\mathrm{r}}$ system, the IAM preambles with guard symbols require $2N_{\mathrm{t}}+1$ OFDM/OQAM symbols. This is equivalent to $N_{\mathrm{t}}+1/2$ complex OFDM/QAM symbols, 
that is, half a complex symbol longer than the preamble one would use in MIMO-OFDM. Nevertheless, even this extra OQAM symbol can be avoided in practice, by eliminating the leading null symbol in the preamble above. This is often possible when interference from preceding data can be practically ignored, as for example in WiMAX~\cite{agm07}, where the gap between two successive frames can play the role of the missing guard symbol. The number of complex OFDM symbols spent for the E-IAM-C preamble is somewhat larger: $N_{\mathrm{t}}/2$ more than that one would use for MIMO-OFDM. However, as it is demonstrated in the simulation examples, this 
increase in the preamble duration can offer considerable improvement in the estimation accuracy. 
\item Eq.~\pref{eq:y1y3} relies on the assumption that the prototype filter is sufficiently well localized in
time so that interference from time instants other than the previous and next is negligible. In fact, often this is not the case in practice and there is nonnegligible interference to the $(p,q)$ FT point
from time instants $q\pm 2$ as well. Hence the ``pseudo pilot" matrix in~\pref{eq:y1y3} is equal to 
\[
\bmx{cr} c^{1}_{p} & c^{1}_{p}\\ c^{2}_{p} & -c^{2}_{p} \emx
\]
instead, with $c^{i}_{p}$ corresponding to the $i$th transmit antenna and $|c^{1}_{p}|\neq |c^{2}_{p}|$. The latter implies that the two columns of the CFR matrix (i.e., the channels from the two transmit antennas) are not estimated with the same accuracy. The associated MSE for the subcarrier $p$ is then equal to $\frac{N_{\mathrm{r}}\sigma^{2}}{2}\left(\frac{1}{|c^{1}_{p}|^{2}}+\frac{1}{|c^{2}_{p}|^{2}}\right)$.
Things are analogous, yet somewhat more complicated, for $N_{\mathrm{t}}>2$. In general, the pseudo pilot matrix is no longer unitary.
It only has (nearly) orthogonal rows. 
This fact was taken into account in the simulations; that is, the correct (not the ideal) pseudo pilot matrix was used in all cases. 
Note that this difficulty can be easily overcome, albeit at an extra cost in training bandwidth, if more than one guard symbol are placed among the $N_{\mathrm{t}}$ repetitions of the IAM preamble above.
\end{enumerate}

\subsection{Optimal sparse preamble}
\label{subsec:Wu}

The MIMO preambles described above are of duration proportional to the number of transmit antennas, $N_{\mathrm{t}}$. Shorter preambles, consisting of 3 OFDM/OQAM symbols per antenna (independently of $N_{\mathrm{t}}$) as in the SISO case (with
zero guard symbols surrounding a nonzero one), were also reported in~\cite{d4.1} and are of the \emph{sparse} type, i.e. with nulls among the pilot subcarriers. One of those preambles was based on earlier work on \emph{orthogonal} training design for MIMO-OFDM systems~\cite{whg05} and shown to be also optimal in the MSE sense. The idea is briefly recalled here. Such a channel estimation approach was also recently adopted in~\cite{cp12}.

Let $N=\frac{M}{L_{h}}$ be an integer,\footnote{If not, one can use for $L_{h}$ the smallest power of~2 that is not smaller than $L_{h}$, that is, $2^{\lfloor \log_{2}L_{h}\rfloor}$.}
greater than or equal to $2N_{\mathrm{t}}$.\footnote{$N\geq N_{\mathrm{t}}$ anyway, since the number of unknown parameters ($N_{\mathrm{t}}N_{\mathrm{r}}L_{h}$) cannot exceed the number of received
signals ($MN_{\mathrm{r}}$).} Consider $N_{\mathrm{t}}$ sets of $L_{h}$ pilots each, placed at subcarriers $\{p_{i},p_{i}+N,p_{i}+2N,\ldots,p_{i}+(L_{h}-1)N\}$, $i=1,2,\ldots,N_{\mathrm{t}}$, with user-chosen starting positions $p_{1},p_{2},\ldots,p_{N_{\mathrm{t}}}\in\{0,1,\ldots,N-1\}$ and with the rest of the subcarriers being nulled. Thus, the pilot subcarriers within each set are equispaced. Moreover, the pilot symbols are chosen to be equipowered. $p_{i}$'s can be so selected as to have at least one zero between two nonzero pilots (sparse preamble), thus avoiding interference among them. 

Since the focus will be on the middle nonzero preamble vector in each antenna, the time index (normally $q=1$) will be dropped for simplicity. 
Let
\[
\gss{y}{}{(p)}=\bmx{cccc} \gss{y}{p}{T} & \gss{y}{p+N}{T} & \cdots & \gss{y}{p+(L_{h}-1)N}{T} \emx^{T},
\]
with $\gss{y}{p}{}\equiv \gss{y}{p,1}{}$ defined as in~\pref{eq:OQAMapproxMIMO}, contain the received signals at the
$p$th pilot-tone set and similarly define $\gss{\eta}{}{(p)}$ for the corresponding frequency domain noise components. Then one can readily verify that 
\begin{eqnarray}
\gss{y}{}{(p)} & = & \gss{D}{1}{(p)}\gss{\bar{F}}{L_{h}}{}\gss{W}{}{(p)}\gss{h}{}{\cdot,1}+\gss{D}{2}{(p)}\gss{\bar{F}}{L_{h}}{}\gss{W}{}{(p)}\gss{h}{}{\cdot,2}+\cdots+\gss{D}{N_{\mathrm{t}}}{(p)}\gss{\bar{F}}{L_{h}}{}\gss{W}{}{(p)}\gss{h}{}{\cdot,N_{\mathrm{t}}} + \gss{\eta}{}{(p)} \nonumber \\
& = & \sum_{i=1}^{N_{\mathrm{t}}}\gss{C}{p}{i}\gss{h}{}{\cdot,i}+\gss{\eta}{}{(p)},
\end{eqnarray}
where 
\[
\gss{C}{p}{i}=\gss{D}{i}{(p)}\gss{\bar{F}}{L_{h}}{}\gss{W}{}{(p)},
\]
with
\[
\gss{D}{i}{(p)}=\mathrm{diag}\left(d_{p}^{i},d_{p+N}^{i},\ldots,d_{p+(L_{h}-1)N}^{i}\right)\otimes\gss{I}{N_{\mathrm{r}}}{}, 
\]
\[
\gss{\bar{F}}{L_{h}}{}=\ga{F}{1:N:M,1:L_{h}}\otimes\gss{I}{N_{\mathrm{r}}}{},
\]
with $\ga{F}{1:N:M,1:L_{h}}$ being (in Matlab notation) the submatrix of the $M$th-order DFT matrix $\g{F}$ consisting of its first $L_{h}$ columns and every $N$th of its rows,
\[
\gss{W}{}{(p)}=\mathrm{diag}\left(1,w_{M}^{p},w_{M}^{2p},\ldots,w_{M}^{(L_{h}-1)p}\right)
\]
with $w_{M}=e^{-\jmath\frac{2\pi}{M}}$ denoting the $M$th root of unity, and $\gss{h}{}{\cdot,i}$ contains the impulse responses of the channels from transmit antenna $i$ to all receive antennas. Then one can compute an estimate of the entire system impulse response $\g{h}=\bmx{cccc} (\gss{h}{}{\cdot,1})^{T} & (\gss{h}{}{\cdot,2})^{T} & \cdots & (\gss{h}{}{\cdot,N_{\mathrm{t}}})^{T} \emx^{T}$ by solving the 
$N_{\mathrm{t}}N_{\mathrm{r}}L_{h}\times N_{\mathrm{t}}N_{\mathrm{r}}L_{h}$ system of equations
\[
\underbrace{\bmx{c} \gss{y}{}{(p_{1})} \\ \gss{y}{}{(p_{2})} \\ \vdots \\ \gss{y}{}{(p_{N_{\mathrm{t}}})} \emx}_{\g{y}} = 
\underbrace{\bmx{cccc} \gss{C}{p_{1}}{1} & \gss{C}{p_{1}}{2} & \cdots & \gss{C}{p_{1}}{N_{\mathrm{t}}} \\
\gss{C}{p_{2}}{1} & \gss{C}{p_{2}}{2} & \cdots & \gss{C}{p_{2}}{N_{\mathrm{t}}} \\
\vdots & \vdots & \ddots & \vdots \\
\gss{C}{p_{N_{\mathrm{t}}}}{1} & \gss{C}{p_{N_{\mathrm{t}}}}{2}  & \cdots & \gss{C}{p_{N_{\mathrm{t}}}}{N_{\mathrm{t}}}  \emx}_{\g{C}} \g{h}+\underbrace{\bmx{c} \gss{\eta}{}{(p_{1})} \\ \gss{\eta}{}{(p_{2})} \\ \vdots \\ \gss{\eta}{}{(p_{N_{\mathrm{t}}})}\emx}_{\g{\eta}},
\]
or 
\begin{equation}
\g{y}=\g{C}\g{h}+\g{\eta}
\label{eq:y=Ch+e}
\end{equation}
Note that $\g{\eta}$ is white. 
Thus, the LS solution of~\pref{eq:y=Ch+e} will be optimal in the MSE sense if $\g{C}$ is unitary. A sufficient condition ensuring the latter is that 
\begin{equation}
d_{p}^{i}=d_{p+N}^{i}=\cdots =d_{p+(L_{h}-1)N}^{i}, \mbox{\ \ for all\ }p,i
\label{eq:equald}
\end{equation}
and the matrix 
\begin{equation}
\g{D}=\bmx{cccc} d_{p_{1}}^{1} & d_{p_{1}}^{2} & \cdots & d_{p_{1}}^{N_{\mathrm{t}}} \\ d_{p_{2}}^{1} & d_{p_{2}}^{2} & \cdots & d_{p_{2}}^{N_{\mathrm{t}}} \\
\vdots & \vdots & \ddots & \vdots \\ d_{p_{N_{\mathrm{t}}}}^{1} & d_{p_{N_{\mathrm{t}}}}^{2} & \cdots & d_{p_{N_{\mathrm{t}}}}^{N_{\mathrm{t}}}\emx
\label{eq:Dmatrix}
\end{equation}
be unitary. For example, with $M=8$, $N_{\mathrm{t}}=2$ and $L_{h}=2$, we have $N=4$. Choosing $\g{D}$ to be the orthogonal matrix 
\[
\g{D}=\bmx{cr} 1 & 1 \\ 1 & -1 \emx
\]
and $p_{1}=0,p_{2}=2$, the middle (nonzero) vectors of the preambles employed at the two transmit antennas will be 
\[
\begin{array}{ccr}
1 & & 1 \\
0 & & 0 \\
1 & & -1 \\
0 & & 0 \\
1 & & 1 \\
0 & & 0 \\
1 & & -1 \\
0 & & 0 \\
\end{array}
\]
Note that, in the SISO case, the above reduces to the optimal sparse preamble-based approach of~\cite{kkrt10} as outlined previously. In fact, it can be verified that, in that case, the pilot symbols need only be equipowered (not as in~\pref{eq:equald}) as proved in~\cite{kkrt10}.  

\section{Simulation Results}
\label{sec:sims}

In this section, simulation results are reported that demonstrate and compare the estimation performances of the IAM schemes previously discussed. In the SISO scenario, the iterative 
method of Hu~\emph{et al.}~\cite{hwl09} was also tested, with the preamble built as in~Fig.~\ref{fig:ICMpreamble}(d), preceded by zeros (due, e.g., to the inter-frame gap) and followed by 
random i.i.d.\ data. Results of the POP method were not included since, as expected, it had a quite poor performance compared with the rest of the methods studied. For the sake of completeness, CP-OFDM is also included, with the minimum possible CP length (equal to the channel order). 
A realistic scenario where all the preambles are followed by pseudo-random data\footnote{In fact, what comes immediately after the preamble is usually control information rather than random data (e.g., \cite{agm07}). Nevertheless, whatever structure is found in this control part of the frame is ignored here for the purposes of the experiments.} was considered. This is of importance for 
the results to be presented, since the assumption used above that the middle preamble OFDM/OQAM symbol receives negligible interference from the following data is not exact, even with
the well localized prototype filters employed in the experiments. Thus, the late part of the preamble signal contains a portion of the front tail of the data burst as well, and \emph{this was taken into account when estimating the required transmit power.}
Note that this problem is not present in CP-OFDM and is due to the long tails of the OFDM/OQAM prototype filter impulse response. It should also be pointed out here that the SFB output signal
due to the (SISO) 3-symbol OFDM/OQAM IAM preamble is of length $(K+1)M$, which is approximately $K+1$ times as long as that of CP-OFDM. One could consider shortening the preamble burst by eliminating its tails, as it was recently described in~\cite{b10c}. As for the time invariance assumption for the channel and its validity throughout the preamble signal, one can see that, for scenarios similar to that of WiMAX for example \cite{agm07}, and for reasonable mobile speeds, the channel coherence interval spans many more than $K+1$ $M$-blocks. 

\begin{figure}
\begin{center}
	\includegraphics[width=6in]{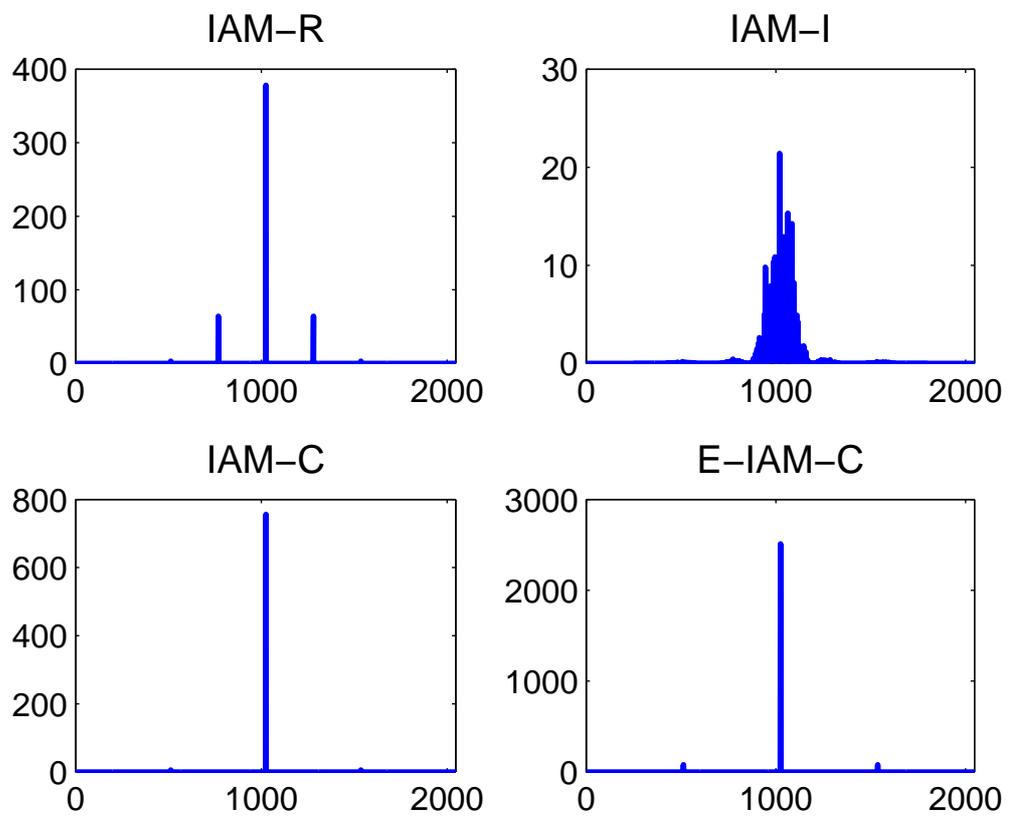}
\end{center}
	\caption{Magnitude squared of the modulated IAM preambles. $M=512$, $K=3$. QPSK modulation.}
	\label{fig:Peaks}
\end{figure}

The experiments utilized prototype filters designed as in~\cite{b01}. Filter banks with $M=512$ and $K=3$ were used. QPSK modulation was adopted. 
The normalized MSE (NMSE), $E\left(\left\|\gss{H}{}{}-\gss{\hat{H}}{}{}\right\|^{2}/\left\|\gss{H}{}{}\right\|^{2}\right)$,
where $\g{H}$ is the CFR and $\g{\hat{H}}$ its estimate, is plotted with respect to the signal to noise ratio (SNR). The latter is defined as the ratio of the channel input (i.e., SFB output) power to the power of the noise. Note that in order to be fair with respect to the power transmission requirements, all the preambles are appropriately scaled so as to result in the \emph{same} power \emph{at the SFB output}. It must be emphasized that these power values turn out to differ much more with each other than they do at the SFB input. (In fact, earlier comparison studies seem not to have 
taken this into account and only equalized powers at the SFB input.) An example is shown in Fig.~\ref{fig:Peaks}. 
The figure also shows that, as pointed out in~\cite{l08}, this kind of IAM preambles result in high PAPR signals at the SFB output, due to their
deterministic/periodic nature.  In fact, as it can be seen in the example, this effect is more evident in IAM-C than in IAM-R, while it is much less severe in IAM-I due to its pseudorandom symbols. E-IAM-C exhibits even higher peaks than IAM-C. These peaks were included in the average power estimates of the corresponding channel inputs. 

\subsection{SISO Systems}
\label{subsec:SISO}

\begin{figure}
\begin{center}
\begin{tabular}{c}
\includegraphics[width=4in]{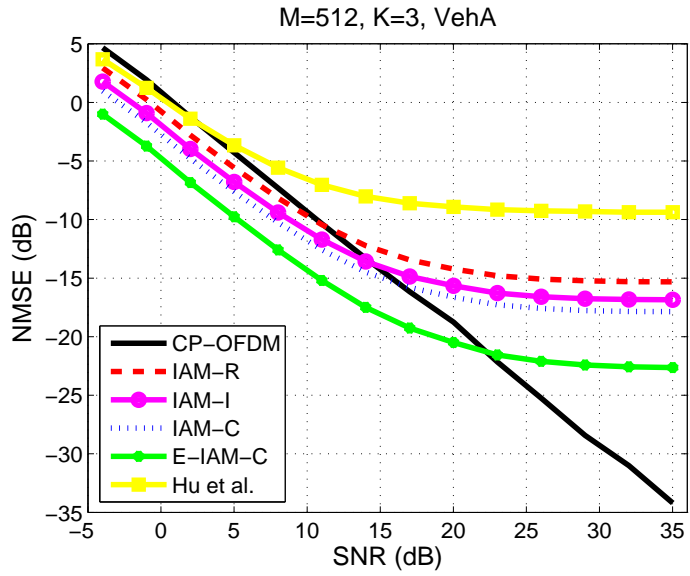} \\
(a) \\
\includegraphics[width=4in]{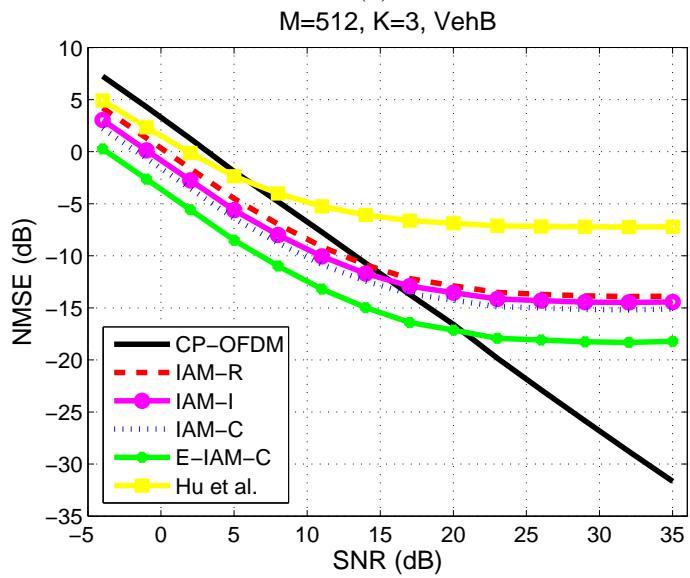} \\
(b) 
\end{tabular}
\end{center}
\caption{Estimation (NMSE) performance of the preamble-based methods for SISO systems. Filter banks with $M=512$, $K=3$ have been used.
Rayleigh channels were assumed, with a (a) Veh-A and (b) Veh-B profile.}
\label{fig:MSE_SISO}
\end{figure}

Training in CP-OFDM was based on a single-symbol complex QPSK pseudo-random preamble. The iterations in the method of~\cite{hwl09} relied on a single data OFDM/OQAM symbol following the pilot symbol.
At moderate to high SNR values, two iterations sufficed for convergence.
Figs.~\ref{fig:MSE_SISO}(a) and (b) show the MSE performance of the methods under study for Veh-A and Veh-B channels~\cite{itu97}, respectively. 
Observe that in both cases E-IAM-C outperforms the other schemes in the whole SNR range considered, with the iterative method of Hu~\emph{et al.} exhibiting the worst performance among the OFDM/OQAM schemes. Moreover, all the OFDM/OQAM methods 
perform better than CP-OFDM at low to moderate SNR values. At higher SNRs, CP-OFDM takes over while the NMSE curves of the OFDM/OQAM schemes reach an error floor. This is a well known phenomenon in such systems and is due to the fact that the approximation~\pref{eq:OQAMapprox} is not exact for channels of significant time dispersion. Hence there is residual intrinsic interference, which is hidden by the noise at low SNRs and shows up in the weak noise regime. It should also be noticed that the OFDM/OQAM methods are less well performing with channels of higher frequency selectivity
and the crossing point with the CP-OFDM curve appears earlier in that case (see Fig.~\ref{fig:MSE_SISO}(b)). This behavior is again explained by the fact that, in the highly frequency selective case, the assumption of a locally flat CFR, and hence~\pref{eq:OQAMapprox}, is much less valid. 

\begin{figure}
\begin{center}
\includegraphics[width=4in]{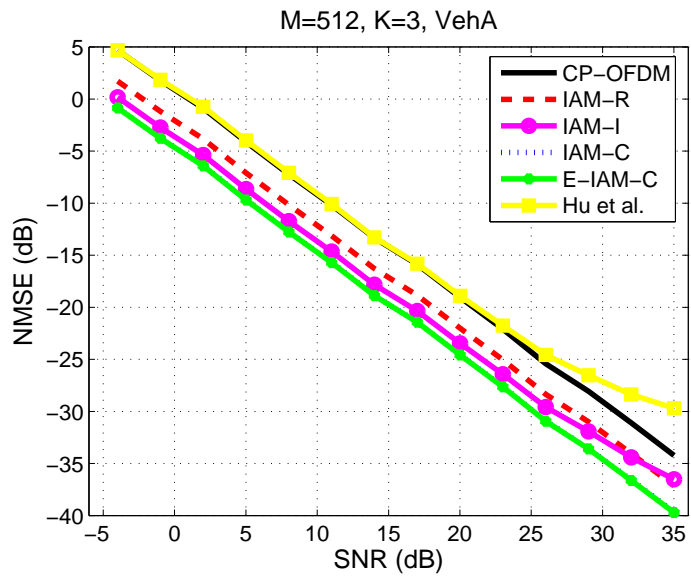} \
\end{center}
\caption{As in~Fig.~\ref{fig:MSE_SISO}(a), but with preamble powers equalized only at the SFB input.}
\label{fig:MSE_SISO_woN}
\end{figure}

Fig.~\ref{fig:MSE_SISO_woN} provides an example of the performances obtained when the preambles are only normalized to equal powers at the SFB \emph{input} (instead of at the transmitted signal as above), as it is the case in most of the existing literature. Comparing with Fig.~\ref{fig:MSE_SISO}(a), one can see that the performance of all methods is then improved (with the error floor effect being significantly remedied) while the differences among them are reduced. 

\subsection{MIMO Systems}
\label{subsec:MIMO}

The channel taps were modeled in accordance with the Kronecker model, that is $\ga{h}{l}=\gssa{R}{\mathrm{r}}{1/2}{l}\gssa{h}{\mathrm{w}}{}{l}\gssa{R}{\mathrm{t}}{1/2}{l}$, $l=0,1,\ldots,L_{h}-1$,
where the receive and transmit correlation matrices $\gssa{R}{\mathrm{t}}{}{l},\gssa{R}{\mathrm{r}}{}{l}$ are generated according to the
exponential model, and $\gssa{h}{\mathrm{w}}{}{l}$ is a matrix of i.i.d.\ zero mean circularly symmetric
Gaussian entries following a Veh-A or Veh-B power delay profile. Weak spatial correlation was assumed at both the transmit and receive sides of the channel.  
The CP-OFDM preamble was similarly chosen, namely $\gss{A}{N_{\mathrm{t}}}{}\otimes \g{x}$ with $\g{x}$ being a pseudorandom $M$-vector of complex QPSK pilots.  

\begin{figure}
\begin{center}
\begin{tabular}{c}
\includegraphics[width=4in]{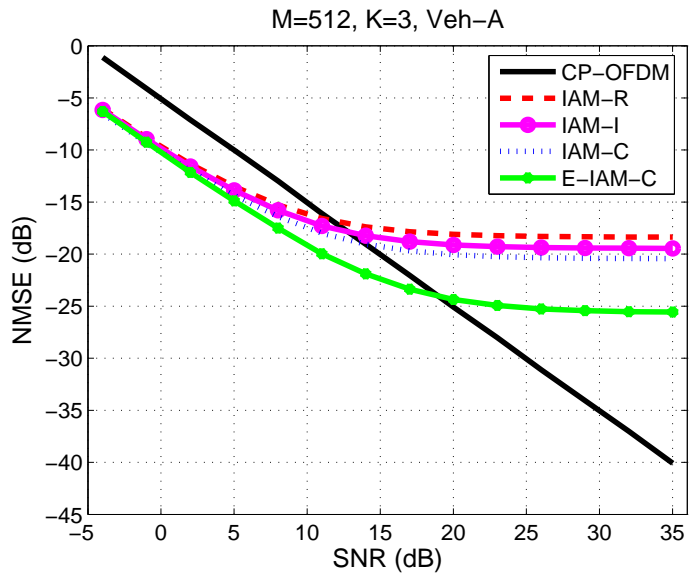} \\
(a) \\
\includegraphics[width=4in]{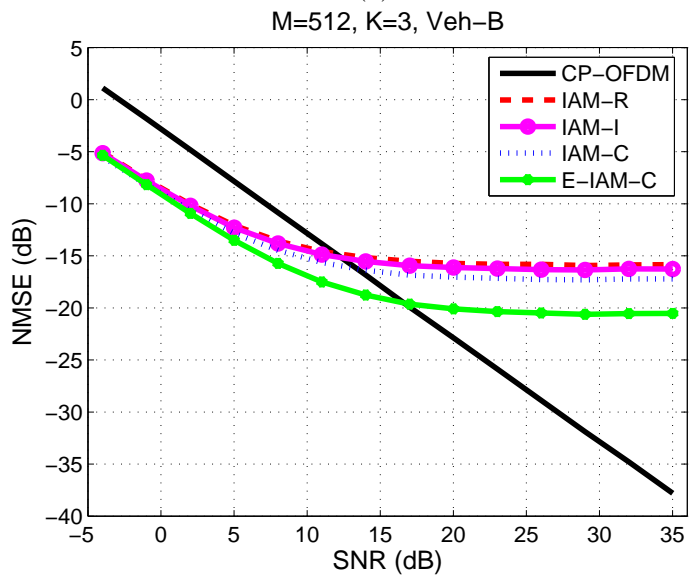} \\
(b) 
\end{tabular}
\end{center}
\caption{Estimation (NMSE) performance of the IAM preamble-based methods for MIMO $2\times 2$ systems. CP-OFDM is included for the sake of comparison. Filter banks with $M=512$, $K=3$ have been used.
Rayleigh channels of (a) Veh-A and (b) Veh-B profiles have been assumed.}
\label{fig:MSE_MIMO}
\end{figure}

Fig.~\ref{fig:MSE_MIMO} shows the MSE performance of the methods under study in $2\times 2$ systems.
Veh-A and Veh-B channels were considered in Figs.~\ref{fig:MSE_MIMO}(a) and (b), respectively. Once more, E-IAM-C outperforms the other schemes in the whole SNR range considered. Moreover, all the IAM methods 
perform better than CP-OFDM at low to moderate SNR values. At higher SNRs, the error floor effect, which was observed in the SISO scenario, appears here too. As previously, it is more
severe for more frequency selective channels (cf.~Fig.~\ref{fig:MSE_MIMO}(b)). 

\section{Conclusions}
\label{sec:concls}

The problem of preamble-based channel estimation in OFDM/OQAM systems was revisited in this paper, and a review of the preamble structures and associated estimation
methods was given, for both SISO and MIMO systems. Results on MSE-optimal design of preambles with zero guard symbols were also reviewed for SISO systems. Analogous results for the MIMO scenario
are still to be derived. 
The estimation performances of the principal methods (including IAM variants and the iterative method of~\cite{hwl09}) were demonstrated via results of simulation in a realistic (including interference from data) and fair scenario and for both mildly and highly frequency-selective channels. A recently developed method of the IAM type, E-IAM-C, was seen to outperform the rest under such conditions and at all considered SNR values. Further investigation is required to assess the loss in performance from tail truncation in the IAM preambles and mitigate the high PAPR problem associated with their periodic structure. Moreover, improved IAM designs as well as optimality results are needed for MIMO-OFDM/OQAM. 

\section{Acknowledgment}
\label{sec:ack}

Part of this work was done within the FP7 project PHYDYAS (Contract no.: ICT-211887; http://www.ict-phydyas.org). D.~Katselis was additionally supported by the European Research Council under the advanced grant LEARN, contract 267381, and by the Swedish Research Council under contract 621-2010-5819.

E.\ Kofidis would like to acknowledge fruitful discussions with Drs. P.~Siohan, C.~L\'{e}l\'{e}, J.~Du, and T.~Hidalgo~Stitz. 

\begin{appendices}

\section{Time-Frequency Neighborhood}

In this appendix, the interference terms from the neighbors of a FT point are computed in closed form, based on the knowledge of the prototype filter. Moreover, their symmetries are demonstrated. 

These are the quantities we have previously denoted as 
\[
\langle g \rangle_{m+p,n+q}^{m,n}=\Im \{ \langle g_{m+p,n+q} | g_{m,n} \rangle \}
\]
where
\[
\langle g_{m+p,n+q} | g_{m,n} \rangle = \sum_{l}g_{m+p,n+q}(l)g_{m,n}^{\star}(l). 
\]
The latter has to be computed for $p\in \{-2,-1,0,1,2\}$ and $q\in\{-1,0,1\}$, with $(p,q)\neq (0,0)$. In general, it can be written as:
\begin{equation}
\langle g_{m+p,n+q} | g_{m,n} \rangle = \sum_{l}g\left(l-(n+q)\frac{M}{2}\right)g\left(l-n\frac{M}{2}\right)e^{\jmath\frac{2\pi}{M}p\left(l-\frac{L_{g}-1}{2}\right)}e^{\jmath\Delta\varphi}
\equiv \jmath \cdot \langle g \rangle_{m+p,n+q}^{m,n}
\label{eq:gg1}
\end{equation}
where $\Delta\varphi\equiv \varphi_{m+p,n+q}-\varphi_{m,n}=(p+q)\frac{\pi}{2}-(mq+pn+pq)\pi$.
Substituting this in \pref{eq:gg1} and after some algebra, the following simplified form results:
\begin{equation}
\langle g_{m+p,n+q} | g_{m,n} \rangle = (-1)^{mq}\jmath^{p+q}e^{-\jmath \frac{2\pi}{M}p\frac{L_{g}-1}{2}}
\sum_{l=\max(-q\frac{M}{2},0)}^{L_{g}-1-\max(q\frac{M}{2},0)}g(l)g\left(l+q\frac{M}{2}\right)e^{\jmath \frac{2\pi}{M}pl}
\label{eq:gg2}
\end{equation}
Observe that this is \emph{independent of $n$}.

Let us consider the various cases in detail: 

1) $q=0,p\in\{\pm1,\pm2\}$

\medskip

\noindent
With $p=\pm 1$, \pref{eq:gg2} becomes
\[
\langle g_{m\pm 1,n} | g_{m,n} \rangle = \pm \jmath e^{\mp \jmath \frac{2\pi}{M}\frac{L_{g}-1}{2}}\sum_{l=0}^{L_{g}-1}g^{2}(l)e^{\pm \jmath \frac{2\pi}{M}l},
\]
and hence 
\begin{equation}
\langle g \rangle_{m\pm 1,n}^{m,n} = \pm e^{\mp \jmath\frac{2\pi}{M}\frac{L_{g}-1}{2}}\sum_{l=0}^{L_{g}-1}g^{2}(l)e^{\pm \jmath\frac{2\pi}{M}l}.
\label{eq:0pm1}
\end{equation} 
In view of the fact that the latter quantities are real and hence equal to their conjugates, we can write
\begin{equation}
\beta \equiv \langle g \rangle_{m+1,n}^{m,n} = -\langle g \rangle_{m-1,n}^{m,n}.
\label{eq:beta}
\end{equation}
For $p=\pm 2$, we have
\[
\langle g_{m\pm 2,n} | g_{m,n} \rangle = -e^{\mp \jmath \frac{2\pi}{M}(L_{g}-1)}\sum_{l=0}^{L_{g}-1}g^{2}(l)e^{\pm \jmath \frac{2\pi}{M}2l}
=-e^{\mp \jmath \frac{2\pi}{M}(L_{g}-1)}G^{(2)}(\pm 2),
\]
with $G^{(2)}(p)$ denoting the $M$-point discrete Fourier transform of $g$. This coincides with the ambiguity function of $g$, $A_{g}(q,p)$, calculated at $(0,\pm 2)$. But the latter is known to equal zero for all $g$'s that can be used as prototype filters in such filter banks ~\cite{ds07b}. Hence, 
\[
\langle g \rangle_{m\pm 2,n}^{m,n}=0
\]

2) $q=-1,p\in\{-2,-1,0,1,2\}$

\medskip

\noindent
With $q=-1$, \pref{eq:gg2} becomes
\[
\langle g_{m+p,n-1} | g_{m,n} \rangle = -\jmath (-1)^{m} \jmath^{p} e^{-\jmath p\frac{2\pi}{M}\frac{L_{g}-1}{2}}\sum_{l=M/2}^{L_{g}-1}g(l)g\left(l-\frac{M}{2}\right)e^{\jmath p\frac{2\pi}{M}l}.
\]
Thus,
\begin{eqnarray}
\langle g \rangle_{m,n-1}^{m,n} & = & -(-1)^{m}\sum_{l=M/2}^{L_{g}-1}g(l)g\left(l-\frac{M}{2}\right) \nonumber \\
& \equiv & -(-1)^{m}\gamma.
\label{eq:0-1}
\end{eqnarray}
For $p=\pm 1$: 
\begin{eqnarray}
\langle g \rangle_{m\pm 1,n-1}^{m,n} & = & -(-1)^{m}\jmath^{\pm 1}e^{\mp \jmath \frac{2\pi}{M}\frac{L_{g}-1}{2}}\sum_{l=M/2}^{L_{g}-1}g(l)g\left(l-\frac{M}{2}\right)e^{\pm \jmath \frac{2\pi}{M}l} \nonumber \\ & \equiv & (-1)^{m}\delta,
\label{eq:pm1-1}
\end{eqnarray}
while for $p=\pm 2$:
\begin{eqnarray}
\langle g \rangle_{m\pm 2,n-1}^{m,n} & = & (-1)^{m}e^{\mp\jmath \frac{2\pi}{M}(L_{g}-1)}\sum_{l=M/2}^{L_{g}-1}g(l)g\left(l-\frac{M}{2}\right)e^{\pm \jmath 2 \frac{2\pi}{M}l} \nonumber \\
& = & (-1)^{m}\epsilon.
\label{eq:pm2-1}
\end{eqnarray}

3) $q=1,p\in\{-2,-1,0,1,2\}$

\medskip

\noindent
Substituting $q=1$ in \pref{eq:gg2} yields
\[
\langle g_{m+p,n+1} | g_{m,n} \rangle = \jmath \cdot \jmath^{p}(-1)^{m}e^{-\jmath p\frac{2\pi}{M}\frac{L_{g}-1}{2}}\sum_{l=0}^{L_{g}-1-\frac{M}{2}}g(l)g\left(l+\frac{M}{2}\right)e^{\jmath p\frac{2\pi}{M}l}
\]
and after a change of variable in the summation:
\[
\langle g_{m+p,n+1} | g_{m,n} \rangle = \jmath \cdot \jmath^{p}(-1)^{m+p}e^{-\jmath p\frac{2\pi}{M}\frac{L_{g}-1}{2}}\sum_{l=\frac{M}{2}}^{L_{g}-1}g(l)g\left(l-\frac{M}{2}\right)e^{\jmath p\frac{2\pi}{M}l}.
\]
Thus,
\begin{eqnarray*}
\langle g \rangle_{m,n+1}^{m,n} & = & (-1)^{m}\sum_{l=\frac{M}{2}}^{L_{g}-1}g(l)g\left(l-\frac{M}{2}\right)
\\
\langle g \rangle_{m\pm 1,n+1}^{m,n} & = & -\jmath^{\pm 1}(-1)^{m}e^{\mp \jmath\frac{2\pi}{M}\frac{L_{g}-1}{2}}\sum_{l=M/2}^{L_{g}-1}g(l)g\left(l-\frac{M}{2}\right)e^{\pm \jmath\frac{2\pi}{M}l}
\\
\langle g \rangle_{m\pm 2,n+1}^{m,n} & = & -(-1)^{m}e^{\mp \jmath\frac{2\pi}{M}(L_{g}-1)}\sum_{l=M/2}^{L_{g}-1}g(l)g\left(l-\frac{M}{2}\right)e^{\pm \jmath 2 \frac{2\pi}{M}l}
\end{eqnarray*}
which, in view of the above, result in:
\begin{eqnarray}
\langle g \rangle_{m,n+1}^{m,n} & = & (-1)^{m}\gamma,
\label{eq:01} \\
\langle g \rangle_{m\pm 1,n+1}^{m,n} & = & (-1)^{m}\delta, 
\label{eq:pm10} \\
\langle g \rangle_{m\pm 2,n+1}^{m,n} & = & -(-1)^{m}\epsilon.
\label{eq:pm20}
\end{eqnarray}

It thus turns out that the $5\times 3$ neighborhood of the $(m,n)$ point is as follows:
\begin{equation}
\begin{array}{rrr}
(-1)^{m}\epsilon & 0 & -(-1)^{m}\epsilon \\
& & \\
(-1)^{m}\delta & -\beta & (-1)^{m}\delta \\
& & \\
-(-1)^{m}\gamma & d_{m,n} & (-1)^{m}\gamma \\
& & \\
(-1)^{m}\delta & \beta & (-1)^{m}\delta \\
& & \\
(-1)^{m}\epsilon & 0 & -(-1)^{m}\epsilon 
\end{array}
\label{eq:pattern_m}
\end{equation}

\end{appendices}

\newpage

\newpage ~ \newpage

\newpage ~ \newpage
\begin{figure}
	\centerline{\begin{tabular}{ccccccccc}
$\begin{array}{crc} 0 & 1 & 0 \\ 0 & -1 & 0 \\ 0 & -1 & 0 \\ 0 & 1 & 0 \\ 0 & 1 & 0 \\ 0 & -1 & 0 \\ 0 & -1 & 0 \\ 0 & 1 & 0 \end{array}$ & & 
$\begin{array}{crc} 0 & d_{0} & 0 \\ 0 & -\jmath d_{0} & 0 \\ 0 & -d_{0} & 0 \\ 0 & -d_{1} & 0 \\ 0 & \jmath d_{1} & 0 \\ 0 & d_{1} & 0 \\ 0 & -d_{0} & 0 \\ 0 & \jmath d_{0} & 0 \end{array}$ & & 
$\begin{array}{crc} 0 & 1 & 0 \\ 0 & -\jmath & 0 \\ 0 & -1 & 0 \\ 0 & \jmath & 0 \\ 0 & 1 & 0 \\ 0 & -\jmath & 0 \\ 0 & -1 & 0 \\ 0 & \jmath & 0 \end{array}$
& & 
$\begin{array}{rrr} \jmath & 1 & -\jmath \\ -1 & -\jmath & 1 \\ -\jmath & -1 & \jmath \\ 1 & \jmath & -1 \\ \jmath & 1 & -\jmath \\ -1 & -\jmath & 1 \\ -\jmath & -1 & \jmath \\ 1 & \jmath & -1 \end{array}$ & & 
$\begin{array}{rrr} -1 & 1 & 1 \\ -\jmath & -\jmath & \jmath \\ -1 & -1 & 1 \\ -\jmath & \jmath & \jmath \\ -1 & 1 & 1 \\ -\jmath & -\jmath & \jmath \\ -1 & -1 & 1 \\ -\jmath & \jmath & \jmath \end{array}$ 
\\ \\
(a) & & (b) & & (c) & & (d) & & (e) 
\end{tabular}}
\caption{Preamble structures for (a) IAM-R , (b) IAM-I, (c) IAM-C methods, and E-IAM-C method for (d) $\epsilon\geq 0$ and (e) $\epsilon<0$. $M=8$. OQPSK modulation is assumed. $d_{0},d_{1}$ are randomly chosen BPSK symbols.}
\label{fig:IAMSISO}
\end{figure}

\newpage ~ \newpage
\begin{figure}
\centering
\begin{tabular}{cccc}
$\begin{array}{rrr} 0 & 1 & 0 \\ 0 & 0 & 0 \\ 0 & -1 & 0 \\ 0 & 0 & 0 \\ 0 & 1 & 0 \\ 0 & 0 & 0 \\ 0 & -1 & 0 \\ 0 & 0 & 0 \end{array}$ & 
$\begin{array}{rrr} 0 & 1 & 0 \\ 0 & -1 & 0 \\ 0 & 1 & 0 \\ 0 & -1 & 0 \\ 0 & 1 & 0 \\ 0 & -1 & 0 \\ 0 & 1 & 0 \\ 0 & -1 & 0 \end{array}$ & 
$\begin{array}{rrr} 1 & 1 & 1 \\ -1 & -1 & -1 \\ -1 & 1 & -1 \\ 1 & -1 & 1 \\ 1 & 1 & 1 \\ -1 & -1 & -1 \\ -1 & 1 & -1 \\ 1 & -1 & 1 \end{array}$ & 
$\begin{array}{r} 1 \\ -1 \\ 1 \\ -1 \\ 1 \\ -1 \\ 1 \\ -1 \end{array}$ \\
(a) & (b) & (c) & (d) 
\end{tabular}
\caption{Preamble structures aiming at interference cancellation: (a) \cite{hwlxl10}; (b) \cite{kc07}; (c) \cite{yihc08}, and avoidance: (d) \cite{hwl09,dhcl10}. $M=8$. OQPSK modulation is assumed.}
\label{fig:ICMpreamble}
\end{figure}

\newpage ~ \newpage
\begin{figure}
	\centerline{\begin{tabular}{ccc}
$\begin{array}{crcrc} 0 & 1 & 0 & 1 & 0 \\ 0 & -\jmath & 0 & -\jmath & 0 \\ 0 & -1 & 0 & -1 & 0 \\ 0 & \jmath & 0 & \jmath & 0 \\ 0 & 1 & 0 & 1 & 0 \\ 0 & -\jmath & 0 & -\jmath & 0 \\ 0 & -1 & 0 & -1 & 0 \\ 0 & \jmath & 0 & \jmath & 0 \end{array}$ & & 
$\begin{array}{crcrc} 0 & 1 & 0 & -1 & 0 \\ 0 & -\jmath & 0 & \jmath & 0 \\ 0 & -1 & 0 & 1 & 0 \\ 0 & \jmath & 0 & -\jmath & 0 \\ 0 & 1 & 0 & -1 & 0 \\ 0 & -\jmath & 0 & \jmath & 0 \\ 0 & -1 & 0 & 1 & 0 \\ 0 & \jmath & 0 & -\jmath & 0 \end{array}$ \\ 
(a) & & (b)
\end{tabular}}
\caption{IAM-C preamble for a $2\times X$ system, with (a) and (b) correspondig to the two transmit antennas. $M=8$. OQPSK modulation is assumed.}
\label{fig:IAMMIMO}
\end{figure}

\newpage ~ \newpage

\newpage

\newpage ~ \newpage

\newpage ~ \newpage

\end{document}